\documentclass[apj]{emulateapj}
\usepackage{mathptmx}

\def\gtorder{\mathrel{\raise.3ex\hbox{$>$}\mkern-14mu
             \lower0.6ex\hbox{$\sim$}}}
\def\ltorder{\mathrel{\raise.3ex\hbox{$<$}\mkern-14mu
             \lower0.6ex\hbox{$\sim$}}}

\slugcomment{Draft of \today}

\shorttitle{All-sky photometric calibration}

\shortauthors{Ofek}

\begin{document}

\title{Calibrated $griz$ magnitudes of Tycho stars: All-sky photometric calibration using bright stars}
\author{
Eran~O.~Ofek\altaffilmark{1}
}
\altaffiltext{1}{Division of Physics, Mathematics and Astronomy, California Institute of Technology, Pasadena, CA 91125, USA}

\begin{abstract}

Photometric calibration to $\sim5\%$ accuracy 
is frequently needed at arbitrary celestial locations;
however, existing all-sky astronomical catalogs do not
reach this accuracy and time consuming photometric calibration procedures are
required.
I fit the Hipparcos $B_{T}$, and $V_{T}$ magnitudes,
along with the 2MASS $J$, $H$, and $K$ magnitudes
of Tycho-2 catalog-stars with
stellar spectral templates.
From the best fit spectral template derived for each star,
I calculate the synthetic
SDSS $griz$ magnitudes
and constructed an all-sky catalog of $griz$ magnitudes
for bright stars ($V\ltorder12$).
Testing this method on
SDSS photometric telescope observations,
I find that the photometric accuracy, for a single star,
is usually about $0.12$, $0.12$, $0.10$ and $0.08$\,mag
(1\,$\sigma$), for the $g$, $r$, $i$, and $z$-bands, respectively.
However, by using $\sim10$ such stars,
the typical errors per calibrated field (systematic + statistical)
can be reduced to about 0.04, 0.03, 0.02, and 0.02\,mag,
in the $g$, $r$, $i$, and $z$-bands, respectively.
Therefore, in cases for which several calibration stars can be
observed in the field of view of an instrument,
accurate photometric calibration is possible.

\end{abstract}


\section{Introduction}
\label{Introduction}

Often in astronomical research,
absolute photometric accuracy
better than $10\%$, is required.
In many cases, the method of choice is to observe
photometric standards (e.g., Landolt 1992).
However, this requires photometric observing conditions,
and additional observations.
The Sloan Digital Sky Survey (SDSS; York et al. 2000)
provides an excellent photometric calibration
in $ugriz$ bands, with accuracy better than $2\%$ (Adelman-McCarthy et al. 2008).
However, this is available only for about a quarter of
the sky.
Other all-sky visible-light catalogue, like the
USNO-B1 (Monet et al. 2003)
and the US Naval Observatory CCD astrograph catalogue (Zacharias et al. 2004)
provide relatively poor photometric accuracy.
The magnitudes of individual stars in these catalogue are
accurate to about $0.3$\,mag, and even with a large number of
stars there is still a considerable field-to-field 
systematic errors.

In this paper, I calculate the
%Jhonson-Cousins $BVRI$ and
SDSS $griz$ magnitudes of Tycho-2 stars over the entire
sky. In case $\gtorder10$ of these stars are visible in
a camera field of view,
these stars can be used to photometrically calibrate
an astronomical image to accuracy of better than $0.04$\,mag.
The only overhead is that typically a shorter exposure
in which the Tycho-2 catalog stars will not be saturated
is required.

\section{Construction of the catalog}
\label{Cat}

From the Tycho-2 catalog (Hog et al. 2000), I selected 
all stars with $B_{T}<13$~mag and $V_{T}<12$~mag.
This choice was made
in order to remove stars with magnitude uncertainties
$\gtorder0.1$~mag.
I cross correlated this list with the 2MASS catalog (Skrutskie et al. 2006),
and selected only stars which have a single 2MASS
match within $6''$.
The final list contains 1,560,980 stars.
Next, I fitted the Tycho-2/2MASS magnitudes of these stars
with synthetic photometry, in the $B_{T}V_{T}JHK$ bands,
of 131 stellar spectral templates (Pickles 1998).
The synthetic photometry was calculated using the code
of Poznanski et al. (2002).
For each star I preformed two types of fits:
(i) a $\chi^{2}$ fit; (ii) a least square, of residuals, fit
ignoring the Tycho-2 and 2MASS photometric errors.
I note that the spectral templates were not corrected for
Galactic extinction.

A table listing the 1.56 million stars, along
with their coordinates, observed $B_{T}V_{T}JHK$ magnitudes,
fitted $griz$ magnitudes for both kinds of fits,
and corresponding root-mean-square (RMS) and $\chi^{2}$,
is available in the electronic version of this paper,
and via the VizieR
service\footnote{http://webviz.u-strasbg.fr/viz-bin/VizieR. Temporarily the catalog is also available from: ftp://agn.caltech.edu/pub/eran/Tycho/}.
In Table~\ref{Tab-CatDesc} I describe the content of this Table.
The third column in the catalog is a flag indicating
if the star is recommended for use as a standard star (i.e., {\tt Flag}$=1$).
These stars satisfy all of the following criteria:
$J>6$\,mag and $H>6$\,mag and $K>6$\,mag;
error($B_{T}$)$<0.15$\,mag;
error($V_{T}$)$<0.15$\,mag;
and RMS of the best RMS-fit template smaller than 0.15\,mag.
There are 992,235 stars with {\tt Flag}$=1$.

\begin{deluxetable}{lll}
\tablecolumns{3}
\tablewidth{0pt}
\tablecaption{Header description for the catalog of SDSS magnitudes for Tycho-2 stars}
\tablehead{
\colhead{Column}              &
\colhead{Explanations}        &
\colhead{Units}               
}
\startdata
1   & Right Ascension J2000.0 & deg \\
2   & Declination J2000.0     & deg \\
3   & Flag for good standards\tablenotemark{a} & \\
4   & $B_{T}$ magnitude       & mag \\
5   & $B_{T}$ magnitude error & mag \\
6   & $V_{T}$ magnitude       & mag \\
7   & $V_{T}$ magnitude error & mag \\
8   & $J$ magnitude           & mag \\
9   & $J$ magnitude error     & mag \\
10  & $H$ magnitude           & mag \\
11  & $H$ magnitude error     & mag \\
12  & $K$ magnitude           & mag \\
13  & $K$ magnitude error     & mag \\
14  & Best rms-fit template\tablenotemark{b} & \\
15  & rms for the best rms-fit template & mag \\
16  & rms-fit $g$-band magnitude\tablenotemark{c} & mag \\
17  & rms-fit $r$-band magnitude\tablenotemark{d} & mag \\
18  & rms-fit $i$-band magnitude\tablenotemark{e} & mag \\
19  & rms-fit $z$-band magnitude\tablenotemark{f} & mag \\
20  & Best $\chi^{2}$-fit template\tablenotemark{b} & \\
21  & $\chi^{2}$ for the best $\chi^{2}$-fit template (d.o.f.=4) & \\
22  & $\chi^{2}$-fit $g$-band magnitude\tablenotemark{g} & \\
23  & $\chi^{2}$-fit $r$-band magnitude\tablenotemark{h} & \\
24  & $\chi^{2}$-fit $i$-band magnitude\tablenotemark{i} & \\
25  & $\chi^{2}$-fit $z$-band magnitude\tablenotemark{j} & 
\enddata
\tablenotetext{a}{Flag indicating if the star is a good standard. i.e., $J>6$\,mag and $H>6$\,mag and $K>6$\,mag and err($B_{T}$)$<0.15$\,mag and err($V_{T}$)$<0.15$\,mag and minimum RMS$<0.15$\,mag.}
\tablenotetext{b}{Template names adopted from the stellar spectra catalog of Pickles (1998).}
\tablenotetext{c}{A 0.06\,mag was already added to this column.}
\tablenotetext{d}{A 0.04\,mag was already added to this column.}
\tablenotetext{e}{A 0.03\,mag was already added to this column.}
\tablenotetext{f}{A 0.03\,mag was already added to this column.}
\tablenotetext{g}{A 0.07\,mag was already added to this column.}
\tablenotetext{h}{A 0.04\,mag was already added to this column.}
\tablenotetext{i}{A 0.04\,mag was already added to this column.}
\tablenotetext{j}{A 0.03\,mag was already added to this column.}
\tablecomments{List of columns in the
$griz$ synthetic magnitudes catalog of Tycho-2 stars.
The catalog is available
from the electronic version of this paper and from the VizieR service.
}
\label{Tab-CatDesc}
\end{deluxetable}

\begin{figure}
\centerline{\includegraphics[width=8.5cm]{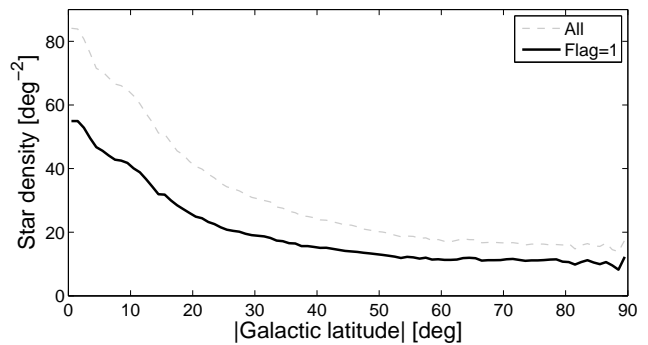}}
\caption{The star density in the constructed catalog,
as a function of Galactic latitude.
The gray-dashed line shows all 1.56 million stars, while
the black line shows the 992,235 stars for which {\tt Flag}$=1$
(i.e., good for photometric calibration).
\label{fig:Tyc2_StarDen}}
\end{figure}

The constructed catalog 
star density as a function of Galactic latitude
is shown in Fig.~\ref{fig:Tyc2_StarDen}.
For {\tt Flag}$=1$ stars, the sky density ranges from 55\,deg$^{-2}$
at the Galactic plane to about 10\,deg$^{-2}$ at the Galactic poles.

\section{Discussion}
\label{Disc}

In order to test the accuracy of the derived synthetic magnitudes,
I constructed a catalog of all the photometric measurements
available from the
SDSS photometric telescope ``secondary patches''
fields\footnote{Available from: http://das.sdss.org/PT/}
(Tucker et al. 2006).
From this catalog I selected all the non-saturated stars
brighter than 12th magnitude in both the $g$- and $r$-bands.
Davenport et al. (2007) analyzed the systematic offset
between the SDSS magnitude system and the SDSS photometric
telescope bands. They found that the magnitudes of very red stars
are different in the two systems.
Using the transformations given by
Davenport et al. (2007) I converted all
the SDSS photometric-telescope magnitudes to the SDSS
system.

\begin{figure*}
\centerline{\includegraphics[width=16.0cm]{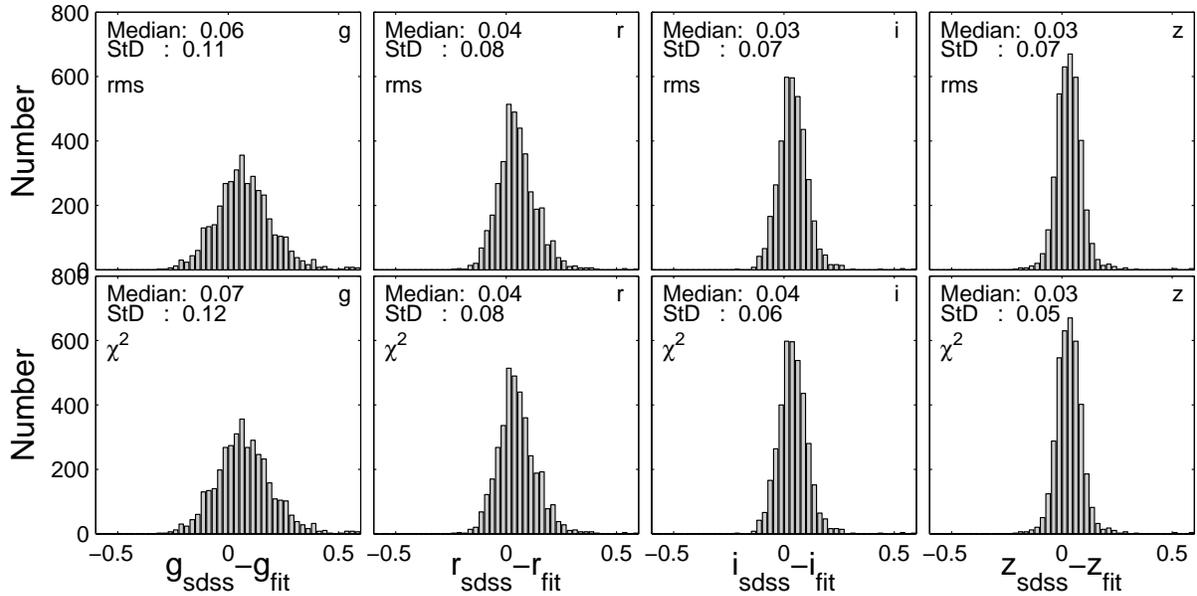}}
\caption{Histograms of the actual SDSS magnitude minus the best fit magnitude
for a set of 3714 stars which magnitude was measured
by the SDSS photometric telescope.
The histograms are shown for the $g$, $r$, $i$, and $z$-bands
(from left to right), respectively.
The upper row is for the best-fit magnitude obtained by minimizing the RMS,
while the lower row is for the $\chi^{2}$ fit, which
is somewhat better.
The median value is the median of the histogram,
while std is half of the range containing $68\%$
of the values in the histogram (rather than the usual definition).
\label{fig:Mag_SDSS_vs_Fit}}
\end{figure*}

Next I cross-correlated this list with the catalog
of $griz$ synthetic magnitudes presented in \S\ref{Cat},
and selected stars which
have $JHK$ magnitudes fainter than 6 (i.e., not saturated);
$B_{T}$ and $V_{T}$ magnitude errors smaller than 0.15;
and RMS of the best RMS-fit template
of less than 0.15\,mag.
For each of the 3714 stars satisfying these criteria,
I compared the
corrected magnitudes (i.e., Davenport et al. 2007)
as measured by the SDSS photometric telescope with
its best-fit synthetic magnitude.
In Figure~\ref{fig:Mag_SDSS_vs_Fit},
I show histograms of the fitted synthetic magnitudes minus the SDSS magnitudes
for the $griz$-bands.
The median value, along with half
the range containing $68\%$ of the stars,
are indicated in each panel.
The upper row is for the best-fit magnitude by minimizing the RMS,
while the lower row is for the $\chi^{2}$ fit.
This plot suggests that the $\chi^{2}$ fit is marginally better.
Therefore, it should be preferred over the RMS-fit.
In all the bands there are small offsets,
listed in Fig.~\ref{fig:Mag_SDSS_vs_Fit},
between the SDSS photometric telescope
magnitudes and the derived synthetic magnitudes.
The magnitudes in the catalog (Table~\ref{Tab-CatDesc}) are corrected
for these small offsets.
I note that I repeated this test using 42 SDSS photometric
standards\footnote{Electronic version available from: http://home.fnal.gov/$\sim$dtucker/ugriz/tab08.dat}
(Smith et al. 2002), and found similar results.

The field of view of large format cameras may contain
only several suitable Tycho-2 stars (see Fig.~\ref{fig:Tyc2_StarDen}).
To estimate the uncertainty in magnitude calibration
when using several stars, I have carried out the following
simulations:
I randomly selected $N$ stars, for any $N$ between 2 and 100, out of the
3714 SDSS standard stars that I used for the
comparison of the derived magnitudes with the actual SDSS magnitudes.
For each set of randomly selected $N$ stars I calculated
the median of differences between the
$\chi^{2}$-fitted synthetic magnitudes and the SDSS magnitudes.
Next, I repeated this procedure 10,000 times (for each $N$),
and calculated the standard deviation of the
10,000 median of differences.
\begin{figure}
\centerline{\includegraphics[width=8.5cm]{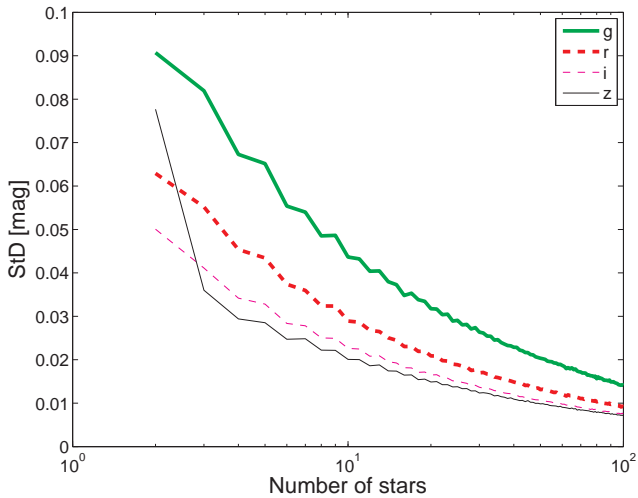}}
\caption{The expected error in photometric calibration as a function
of the number of stars used in the calibration process (see text).
The thick-solid line, thick-dashed line, thin-dashed line,
and thin-solid line, are for the $g$, $r$, $i$, and $z$-bands, respectively.
\label{fig:Error_vs_Nstars}}
\end{figure}
In Figure~\ref{fig:Error_vs_Nstars} I show the
standard deviation of the distribution of the median
of differences
between the synthetic magnitudes
and the measured SDSS magnitudes,
as a function of $N$.
The plot suggests that,
for example, by using five stars the errors in photometry
reduce to
0.07\,mag, 0.04\,mag, 0.03\,mag and 0.03\,mag
in the $g$, $r$, $i$ and $z$-bands, respectively.
When using ten stars the errors reduce to
about 0.04\,mag, 0.03\,mag, 0.02\,mag and 0.02\,mag
in the $g$, $r$, $i$ and $z$-bands, respectively.
%
%Therefore, by using several stars it is possible to reduce the
%errors significantly.
In order to remove outliers it is important to use
the median of differences (i.e., and not mean of differences).

To summarize,
I suggest an alternative method for photometric calibration
that may work to accuracy of about several percents.
This method relies on selected stars in the Tycho-2
catalog for which I fitted spectral templates to
their $B_{T}V_{T}JHK$ magnitudes.
Two major limitation of this method is that several Tycho-2 stars
are needed in the field of view,
and that shorter exposures, in which these stars are not saturated,
are required.
Given the catalog star density as a function
of Galactic latitude (Fig.~\ref{fig:Tyc2_StarDen}),
this method is applicable for instruments
with large field of view or for low Galactic
latitude observations.

\acknowledgments
I would like to thank the referee for constructive comments.
I thank Shri Kulkarni for expressing the need
for an all-sky catalog of standard stars,
and to Avishay Gal-Yam, Dovi Poznanski, Orly Gnat, and Andrew Pickles
for valuable discussions.
This work is supported in part by grants from NSF and NASA.

\end{document}